\documentclass[conference]{IEEEtran}
\IEEEoverridecommandlockouts
\usepackage{cite}
\usepackage{graphicx} 
\usepackage{xurl}
\usepackage{amsmath,amssymb,amsfonts}
\usepackage{textcomp}
\usepackage{hyperref}
\usepackage{listings}
\usepackage{xcolor}
\def\BibTeX{{\rm B\kern-.05em{\sc i\kern-.025em b}\kern-.08em
    T\kern-.1667em\lower.7ex\hbox{E}\kern-.125emX}}

\begin{document}
\title{The Role of AI in Modern Penetration Testing}

\author{\IEEEauthorblockN{J. Alexander Curtis}
\IEEEauthorblockA{\textit{Department of Computer Science} \\
\textit{Boise State University}\\
Boise, ID. USA \\
alexcurtis@u.boisestate.edu}
\and
\IEEEauthorblockN{Nasir U. Eisty}
\IEEEauthorblockA{\textit{Department of EECS} \\
\textit{The University of Tennessee}\\
Knoxville, TN, USA \\
neisty@utk.edu}
}


\maketitle

\begin{abstract}
Penetration testing is a cornerstone of cybersecurity, traditionally driven by manual, time-intensive processes. As systems grow in complexity, there is a pressing need for more scalable and efficient testing methodologies. This systematic literature review examines how Artificial Intelligence (AI) is reshaping penetration testing, analyzing 58 peer-reviewed studies from major academic databases.
Our findings reveal that while AI-assisted pentesting is still in its early stages, notable progress is underway, particularly through Reinforcement Learning (RL), which was the focus of 77\% of the reviewed works. Most research centers on the discovery and exploitation phases of pentesting, where AI shows the greatest promise in automating repetitive tasks, optimizing attack strategies, and improving vulnerability identification.
Real-world applications remain limited but encouraging, including the European Space Agency’s PenBox and various open-source tools. These demonstrate AI’s potential to streamline attack path analysis, analyze complex network topology, and reduce manual workload. However, challenges persist: current models often lack flexibility and are underdeveloped for the reconnaissance and post-exploitation phases of pentesting. Applications involving Large Language Models (LLMs) remain relatively under-researched, pointing to a promising direction for future exploration.
This paper offers a critical overview of AI's current and potential role in penetration testing, providing valuable insights for researchers, practitioners, and organizations aiming to enhance security assessments through advanced automation or looking for gaps in existing research.
\end{abstract}


\section{Introduction}\label{sec:introduction}
Artificial Intelligence (AI) is increasingly permeating cybersecurity, offering opportunities to enhance efficiency and effectiveness across many tasks. One domain ripe for AI-driven innovation is security penetration testing (pentesting), which has traditionally relied on manual labor and expert intuition to uncover system vulnerabilities. As system complexity and potential attack vectors continue to grow, there is an urgent need for scalable, automated solutions.

Penetration testing, commonly referred to as ``pentesting,'' constitutes an essential methodological framework within cybersecurity risk assessment and vulnerability identification. The origins of penetration testing can be traced to the late 1960s and early 1970s, when computer security researchers began developing methodical approaches to test system vulnerabilities. It is a core cybersecurity practice in which ethical hackers simulate real-world attacks to identify security weaknesses.  This security investigation procedure involves the authorization of simulated cyberattacks by experienced and trusted professionals aimed at an organization's own computer systems, networks, and applications. This is done for the purpose of assessing potential security vulnerabilities that could be exploited by malicious entities against these systems. 


Pentesting enables organizations to:

\begin{itemize}
    \item Identify exploitable vulnerabilities before malicious actors do
    \item Validate the effectiveness of existing defenses
    \item Evaluate incident response capabilities under realistic threat scenarios
\end{itemize}

Modern pentesting typically follows a structured, multi-stage approach. The NIST 800-115 framework outlines four key stages~\cite{NIST800115, Shebli2018-ca, Ghanem2019-ow}:

\begin{enumerate}
    \item \textbf{Preparation \& Reconnaissance:} Collect system and network information.
    \item \textbf{Discovery \& Vulnerability Analysis:} Identify potential weaknesses.
    \item \textbf{Exploitation:} Attempt to breach systems using discovered vulnerabilities.
    \item \textbf{Reporting \& Remediation:} Document findings and guide mitigation.
\end{enumerate}

\begin{figure}
    \centering
    \includegraphics[width=0.5\linewidth]{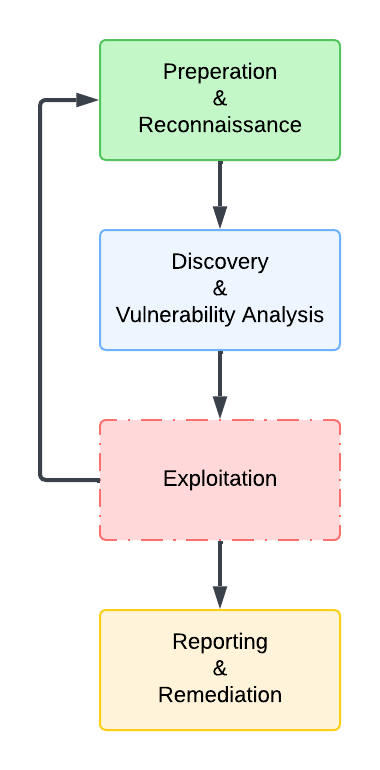}
    \caption{Pentesting Process}
    \label{fig:pentest-process}
\end{figure}

These stages are visualized in Fig.~\ref{fig:pentest-process}, with each stage requiring different tools and expertise. AI has the potential to enhance each step, accelerating reconnaissance, automating the selection of exploit paths, and even assisting in remediation and documentation. We aim to examine how AI contributes across these phases, particularly through the lens of the four research questions outlined in Section~\ref{sec:methodology}.


AI offers promising capabilities for pentesting, by automating repetitive tasks, optimizing attack strategies, and uncovering novel vulnerabilities. Machine Learning (ML), a subset of AI, has already been widely deployed in cybersecurity for tasks such as anomaly detection, security testing, and code analysis. Recent progress in technologies such as RL and LLMs has extended AI’s utility into earlier and more interactive phases of the software development lifecycle, positioning it as a viable force in modern pentesting.

The increasing complexity of digital ecosystems, coupled with the escalating sophistication of cyber threats, have entrenched penetration testing an indispensable component of comprehensive cybersecurity risk management strategies. As technological landscapes continue to evolve, the methodological frameworks and technological tools that support penetration testing will undoubtedly undergo continuous refinement and innovation.

Through a systematic literature review (SLR) of 58 studies, we investigate the current applications, methodologies, and benefits of AI-assisted pentesting, while identifying the challenges and limitations of current approaches. Ultimately, this research seeks to provide a comprehensive understanding of the current state of AI in pentesting and to offer insights into its future potential.

\begin{figure*}
    \centering
    \includegraphics[width=1\linewidth]{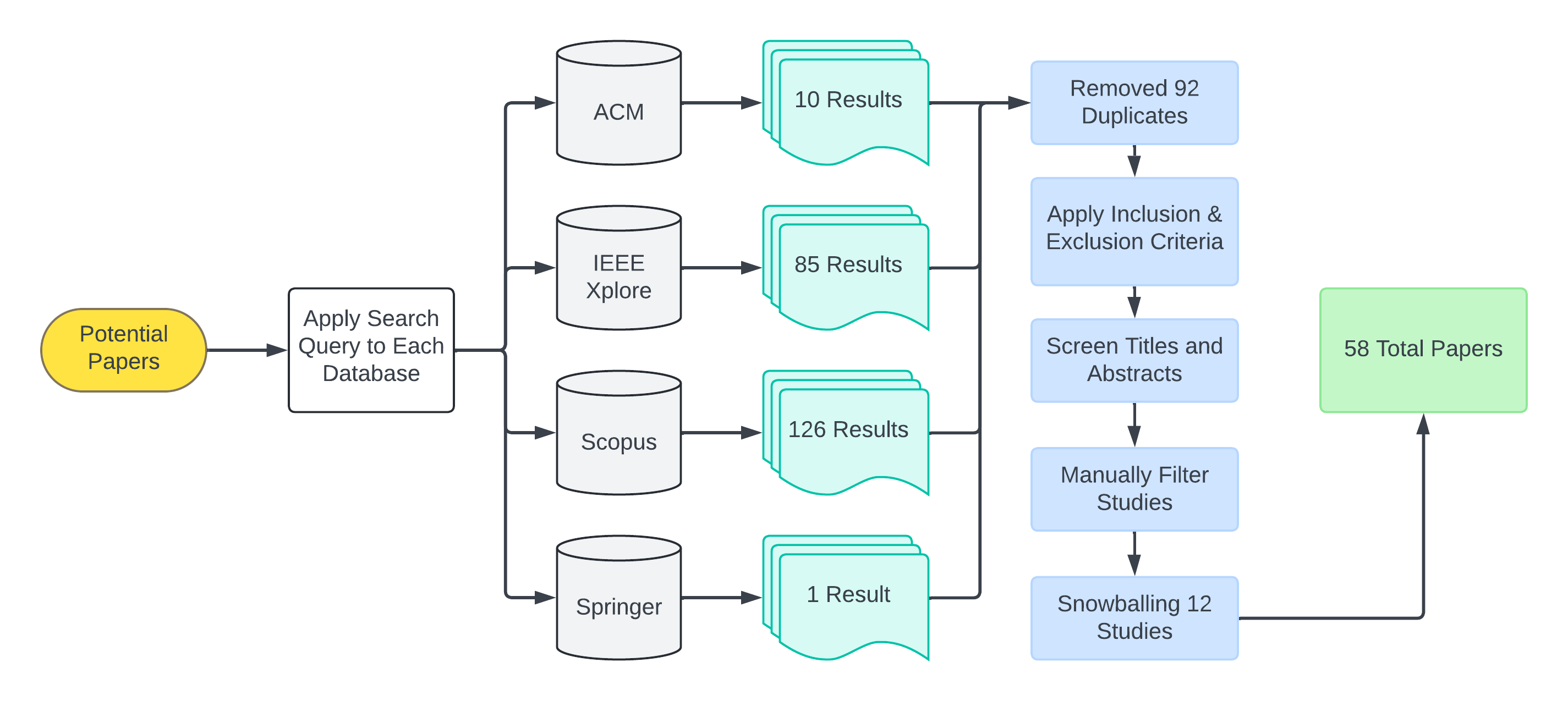}
    \caption{Process Diagram for Paper Selection}
    \label{fig:paper-selection-process}
\end{figure*}

\section{Related Work} \label{sec:relatedwork}
Several prior SLRs have addressed penetration testing, but few focus specifically on AI-assisted pentesting, and none reflect the recent surge in AI advancements.

\subsection{Key Related Studies}

The most directly relevant review is by McKinnel et al.~\cite{McKinnel2019-xs}, which examined AI-assisted pentesting as of 2019. However, it predates critical developments such as LLMs and primarily focuses on RL. The paper itself notes the limited scope of AI techniques explored at the time, leaving a gap in capturing today’s broader AI landscape.


Ghanem and Chen~\cite{Ghanem2019-ow} offer a focused review on using RL to automate pentest tasks and improve coverage. While important, their scope is narrower, rather than offering a broad synthesis of AI methodologies. It is a 2019 publication that also predates recent AI advancements.

Parveen et al.~\cite{Parveen2023-ui} published a recent SLR on pentesting methods in 2023, categorizing various tools and approaches (e.g., mobile, web, client-side), but omitting the role of AI.


Additional SLRs examine penetration testing within specific domains, such as Blockchain~\cite{Kissoon2022-uu, Yulianto2023-gh}, Content Management Systems~\cite{Jagamogan2021-tt}, Industrial Control Systems~\cite{Sindhwad2022-yp, Mohamed2023-qn}, Docker~\cite{Mubanda2023-ha}, and tooling~\cite{Saber2023-bp}—but only mention AI-assisted pentesting as a future direction, not a central focus.


\subsection{Identified Gaps and Motivation for this SLR}

In light of these limitations, our paper aims to fill the gap by presenting a current, comprehensive review of AI-assisted penetration testing. Our goal is to evaluate how modern AI techniques, including RL, LLMs, Deep Learning, and other novel methodologies, are being used to support the penetration testing process in order to identify areas where future work should be focused.




\section{Research Methodology}\label{sec:methodology}
The methodology used for this research is an SLR for the current state of Artificial Intelligence technology to enhance penetration security testing. The systematic methodology of this research follows the Kitchenham and Charters~\cite{Kitchenham2004-nl} methodology outline to carry out an effective SLR in software engineering. 
We follow guidelines for inclusion, exclusion, and a compilation process based on existing literature. Our process is outlined in Fig.~\ref{fig:paper-selection-process} and discussed in Section~\ref{sec:data-collection}.

\subsection{Research Questions}

We seek to answer the following research questions through the course of this research:

\begin{itemize}
    \item{\textbf{RQ1: How has AI been applied to pentesting?}}\\
    This question aims to establish a baseline by examining current applications of AI in penetration testing to date, both in research and industry, helping to contextualize current capabilities and trends.
    \item{\textbf{RQ2: What AI methodologies have been the primary focus of research related to penetration testing?}}\\
    This research question seeks to identify which AI methodologies, such as reinforcement learning, deep learning, and natural language processing, have been explored most prominently in the context of penetration testing. Understanding where research efforts have been concentrated can help highlight promising directions, reveal underexplored areas, and inform future work aimed at improving the efficiency and effectiveness of vulnerability discovery and exploitation.
    \item{\textbf{RQ3: Which phases of the penetration testing process are most likely to benefit from AI assistance?}}\\
    Penetration testing consists of four distinct phases, as outlined in Fig.~\ref{fig:pentest-process}. This research question explores which of these phases stands to benefit most from the application of AI technologies.
    \item{\textbf{RQ4: What are the key benefits and limitations associated with AI-driven penetration testing?}}\\
    This research question aims to critically assess both the advantages and constraints of using AI in penetration testing. By evaluating the practical benefits, such as improved speed, scalability, or accuracy, alongside limitations like interpretability, false positives, or ethical concerns, it helps establish a balanced understanding of AI's value and feasibility in this domain.
\end{itemize}

\begin{table}[ht]
    \centering
    \begin{tabular}{|c|p{5cm}|c|}
    \hline
    Database & Query & Results \\
    \hline
    ACM & (Abstract:(``penetration testing" OR "pentest") AND Abstract:(ai OR ``Artificial Intelligence" OR ``Machine Learning" OR ``Reinforcement Learning")) & 9 \\
    \hline
    IEEE & (((11Abstract":pentest OR ``penetration testing") AND (``Abstract":``AI" OR ``GPT" OR ``LLM" OR ``ML" OR ``Artificial Intelligence" OR ``Machine Learning" OR ``Reinforcement Learning" OR ``RL")) & 126 \\
    \hline
    Scopus & TITLE-ABS-KEY ((("pentest" OR "penetration test" OR "security test") AND AI) AND PUBYEAR \texttt{>} 2020 AND PUBYEAR \texttt{<} 2025 AND (LIMIT-TO (SUBJAREA, ``COMP")) AND (LIMIT-TO (LANGUAGE, ``English"))) & 67 \\
    \hline
    Springer Link & AI Penetration Testing (Conference Paper, Article, Research article) Subdiscipline: Software engineering/programming and operating systems & 19 \\
    \hline
    \end{tabular}
    \caption{Search Queries used for each Database}
    \label{tab:search_queries}
\end{table}

\subsection{Paper Selection}\label{sec:data-collection}

\subsubsection{Search Strategy}

The inclusion of literature for the review started with searching across four major academic research databases: ACM Digital Library, IEEE Xplore, Scopus, and Springer.     \footnotetext{Keywords and abbreviations in search queries that yielded no additional results were removed from the table for brevity, maintaining reproducibility}

\subsubsection{Search Criteria}

The search queries that were used for each database are shown in Table~\ref{tab:search_queries}. While syntax is modified to accomodate each database, the pattern is the same, requiring the mention of ``Penetration Testing" or ``Pentest" in the abstract AND the inclusion of some form of AI word or abbreviation including ``Reinforcement Learning", ``Artificial Intelligence", ``Machine Learning", ``Large Language Model" and their appropriate abbreviations. This approach ensured that each retrieved paper explicitly addressed both the application of AI and its relevance to penetration testing.

\subsubsection{Inclusion and Exclusion Criteria}

After all paper results were compiled, the 82 duplicates were removed, and then we applied inclusion and exclusion criteria to determine which papers were applicable to this review. The inclusion criteria are described in Table~\ref{tab:inclusion-criteria}, and the exclusion criteria are described in Table~\ref{tab:exclusion-criteria}.

\begin{table}[h]
    \centering
    \begin{tabular}{p{1cm}p{7cm}}
        \hline
        \textbf{No.} & \textbf{Inclusion Criteria} \\
        \hline
        IC1 & Full access to the document \\
        IC2 & Must be written in English \\
        IC3 & Must be published in a peer-reviewed journal or conference \\
        IC4 & Must discuss pentesting for purposes of system or network security \\
        IC5 & Must discuss AI involvement (positive or negative) in the process \\
        IC6 & The publication must answer at least one research question \\
        \hline
    \end{tabular}
    \caption{Inclusion Criteria}
    \label{tab:inclusion-criteria}
\end{table}

\begin{table}[h]
    \centering
    \begin{tabular}{p{1cm}p{7cm}}
        \hline
        \textbf{No.} & \textbf{Exclusion Criteria} \\
        \hline
        EC1 & Duplicate studies \\
        EC2 & Penetration Testing in fields outside computer security \\
        EC3 & Publication does not significantly contribute to the area of study \\
        \hline
    \end{tabular}
    \caption{Exclusion Criteria}
    \label{tab:exclusion-criteria}
\end{table}

\subsubsection{Data Analysis}

The final study includes a total of 58 selected papers that matched the criteria and were considered appropriate for this review.

\section{Comparison \& Results} \label{sec:results}

This section of the paper compares and discusses the findings we discovered based on the current literature available from our paper selection process. We address the research questions RQ1 - RQ4 below to guide and focus the comparison analysis. 

\subsection{\textbf{RQ1: Present Use of AI in Pentesting}}

The findings indicate that current practical applications of AI-assisted pentesting remain limited; most uses of AI in this domain are still in the research or proof-of-concept stage. This underscores that AI-assisted penetration testing is still in its infancy, offering significant opportunities for future development and validation.

One notable real-world implementation is by the European Space Agency which developed an AI-driven pentesting platform called \textit{PenBox}~\cite{Happe2023-il, Ghanem2019-ow}. This tool is tailored to detect vulnerabilities early in the development lifecycle~\cite{Happe2023-il}. The tool is limited specifically to space systems, and is optimized for attack patterns unique to that domain. Although its scope is narrow, PenBox demonstrates the considerable promise of AI-assisted penetration testing in operational environments with tangible benefits of cost savings and increased speed of development.

Beyond this single real-world example, several open-source and academic tools highlight how AI is being experimentally applied:
\begin{enumerate}
    \item \textbf{Shennina-based Framework:} Karagiannis et al.~\cite{Karagiannis2024-ns} developed a simulation and validation tool for automated testing built on the Shennina platform.
    \item \textbf{Link:} Lee et al.~\cite{Lee2022-po} proposed a reinforcement learning-based tool to dynamically detect XSS vulnerabilities.
    \item \textbf{Pentraformer:} Wang et al.~\cite{Wang2024-km} introduced a reinforcement learning system for dynamic network discovery that emulates human attacker behavior.
    \item \textbf{SetTron:} Yang et al.~\cite{Yang2023-oe} presented a deep reinforcement learning model to compute efficient penetration paths without prior knowledge of the network topology.
    \item \textbf{ASAP:} Chowdhary et al.~\cite{Chowdhary2020-oz} developed a deep neural network model to derive optimal attack policies over large enterprise networks.
    \item \textbf{DUSC-DQN:} Wang et al.~\cite{Wang2022-di} proposed a reinforcement learning approach incorporating a Greedy-UCB algorithm to improve exploration and outperform human attackers in simulated tests.
\end{enumerate}

A particularly novel application is \textit{PenHeal}, developed by Huang and Zhu~\cite{Huang2023-lu}. Unlike other tools focused on attack or exploration phases, PenHeal employs a two-stage pipeline using LLMs: it first identifies vulnerabilities and then guides system administrators through remediation steps. This is the only tool identified in this SLR which explicitly addresses the final phase of the penetration testing process—\textit{Reporting \& Remediation}.

\begin{figure}
    \centering
    \includegraphics[width=1\linewidth]{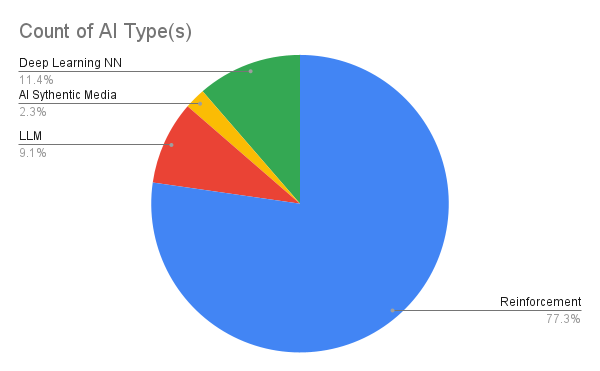}
    \caption{Percentage of Papers Proposing Using Different Types of AI}
    \label{fig:types-of-ai}
\end{figure}

\subsection{\textbf{RQ2: Most Commonly Used AI Methodologies}}

Artificial Intelligence encompasses a wide array of methodologies, including machine learning, reinforcement learning, deep learning, and LLMs, among others. The purpose of this research question is to identify which of these methodologies have been most prominently applied to pentesting in existing literature. This can help guide future research by highlighting where current efforts are concentrated and where opportunities remain for further exploration.

Our analysis reveals a clear dominance of Reinforcement Learning as the AI methodology with the most active research and testing. Of the 54 papers that could be classified by the penetration testing phase\footnote{Four papers were excluded from this categorization due to their broad or conceptual focus. For example, Wang et al.~\cite{Wang2024-zr} discuss the ethical and legal implications of AI in penetration testing rather than any specific technical implementation.}, 42 (approximately 77\%) used reinforcement learning. As shown in Fig.~\ref{fig:types-of-ai}, reinforcement learning accounts for the vast majority of proposed AI applications in pentesting.

A common application of reinforcement learning across these papers is the identification of optimal attack paths in large or complex network environments. These approaches often model an attacker navigating an unknown environment with minimal prior information, simulating intelligent decision-making over time.

Only four papers use LLMs as their AI technology~\cite{Huang2023-lu, Happe2023-il, Naito2023-tm, Gregory2024-pc}. While LLM-based approaches are fewer, they often target different phases of the penetration testing process, such as social engineering, guidance, or remediation.

One notable distinction between RL-based and LLM-based tools lies in their deployment models. Most reinforcement learning tools are designed to run locally, making them attractive for both legitimate testers and malicious actors. In contrast, many LLM-based tools rely on access to public APIs from providers like OpenAI or Anthropic. An exception is the system developed by Gregory and Liao, which uses a locally hosted LLM based on Mistral-7B enhanced with retrieval-augmented generation (RAG) techniques~\cite{Gregory2024-pc}.

Finally, one paper stands out for its unique use of AI synthetic media (AI-SM). Soares et al.~\cite{Soares2023-aj} introduced a method to generate realistic fake identities and documents to improve the social engineering stages of penetration testing. This was the only instance of synthetic media usage among the reviewed studies.

\subsection{\textbf{RQ3: Improvements to Process}}

\begin{figure}
    \centering
    \includegraphics[width=1\linewidth]{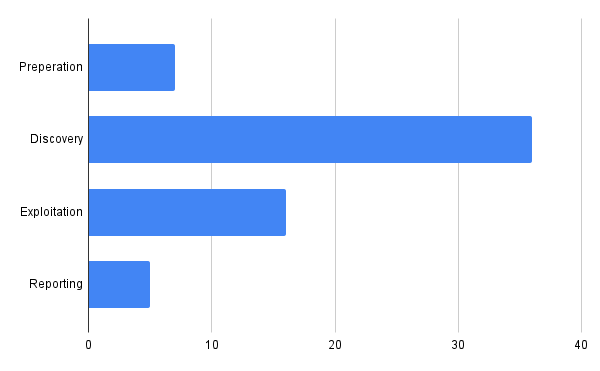}
    \caption{Distribution of Papers Across Phases of the Penetration Testing Process}
    \label{fig:pentest-phase-bar}
\end{figure}

This research question aims to identify which phases of the penetration testing process are most likely to benefit from AI assistance. As described in Section~\ref{sec:introduction} and illustrated in Fig.~\ref{fig:pentest-process}, the process is divided into four primary phases:

\begin{enumerate}
    \item Preparation (and Reconnaissance)
    \item Discovery (and Vulnerability Analysis)
    \item Exploitation
    \item Reporting (and Remediation)
\end{enumerate}

These phases represent broad groupings of tasks within the penetration testing lifecycle. For this literature review, each paper was analyzed and tagged based on the phase(s) it addressed. If a paper contributed to more than one phase, it was tagged accordingly. While some studies covered the full process, the majority focused on a single phase.

The distribution of research contributions across these phases is shown in Fig.~\ref{fig:pentest-phase-bar}. The results clearly show that the majority of AI-focused research targets the \textit{Discovery} phase, followed by the \textit{Exploitation} phase. In contrast, significantly fewer studies address the \textit{Preparation} or \textit{Reporting} phases, with only seven and five papers respectively. This suggests a potential gap in research and highlights areas where AI assistance could be further explored and applied.


It is worth noting that the dominance of the Discovery phase is partially driven by a concentration of studies using reinforcement learning to optimize attack paths in network environments. This has created a dense cluster of research in a single subdomain. However, this does not imply that all areas of the Discovery phase are saturated; many subtopics, such as vulnerability identification beyond network topology, remain underexplored and present opportunities for future work.

\subsection{\textbf{RQ4: Benefits and Limitations for Human/AI Assistance}}

This research question aims to evaluate the benefits and limitations of integrating AI into the penetration testing process. Key potential benefits include increased efficiency, improved accuracy, discovery of novel vulnerabilities, and enhanced accessibility for less experienced testers.

The most frequently cited benefit across the reviewed literature is AI’s ability to accelerate time-consuming aspects of penetration testing~\cite{Gregory2024-pc, Saber2023-bp}. Traditional pentesting is a laborious process that requires engineers to attempt numerous potential attack vectors, most of which fail, before identifying a viable exploit. As modern infrastructure becomes more complex and distributed, the time and expertise required to perform effective tests have grown. Furthermore, network architectures and server systems are more advanced, and there are more resources in the mix that require exploration (load balancers, CDNs, network switches, routers, etc.)~\cite{Van-Hoang2022-ii}. AI-Assisted Pentesting does not attempt to remove human engineers from the process as much as support them by pointing them to potential flaws, saving many hours of attempting vulnerabilities that are dead ends, or identifying attack vectors previously consider infeasible.

For example, Van Hoang et al.~\cite{Van-Hoang2022-ii} developed a reinforcement learning system that observes repetitive tasks performed by human testers and automates them in future tests, thereby reducing repetitive workload. Similarly, Li et al.~\cite{Li2024-ab} introduced an AI agent that suggests optimal attack strategies based on ongoing interactions with human testers, continually refining its recommendations through feedback. Bianou and Batogna~\cite{Bianou2024-lx} created an LLM-based system called \textit{PENTEST-AI}, designed to help novice engineers operate with the effectiveness of seasoned professionals by providing contextual guidance throughout the pentesting process. These examples highlight the broader potential for AI to democratize pentesting expertise and expand workforce capability.

AI has also been used to create realistic simulation environments for training purposes~\cite{Karagiannis2024-ns, Hao2022-uh, Wang2024-lh}. These environments can mimic enterprise network structures and provide safe, repeatable scenarios for engineers to develop and test their skills, ultimately helping to scale pentesting proficiency across teams.

However, the review also uncovered key limitations. One major challenge is the inflexibility of many early AI models, particularly those based on reinforcement learning. These models are often narrowly scoped and must be completely re-trained when new attack types or zero-day vulnerabilities emerge~\cite{Saber2023-bp}, limiting their adaptability in fast-evolving environments.

Another concern is the brute-force nature of many AI scanning techniques. AlMajali et al.~\cite{AlMajali2023-ej} noted that many reinforcement learning-based systems rely on exhaustive exploration, which can generate excessive traffic and alert detection systems. In contrast, human testers tend to use more discreet, strategic probing.

In summary, AI holds great promise for enhancing the speed, scope, and accessibility of penetration testing. However, practical challenges, particularly model adaptability and operational subtlety, must be addressed before widespread deployment can be achieved.

\section{Future Work} \label{sec:futurework}
While this review provides a comprehensive snapshot of current AI-assisted pentesting research, the field remains nascent with significant opportunities for advancement. Two primary directions for future work are outlined below.

\subsubsection{Expanding Coverage Across Pentesting Stages}

Most research focuses on the \textit{Discovery} and \textit{Exploitation} stages, leaving \textit{Preparation} and \textit{Reporting} underexplored. Future efforts should broaden the role of AI throughout the entire lifecycle.

\begin{itemize}
    \item \textbf{Intelligent Reconnaissance:} Use NLP and visual analysis to enhance early-stage system profiling.
    \item \textbf{Automated Reporting and Remediation:} Expand tools like PenHeal~\cite{Huang2023-lu} to generate context-aware, actionable remediation guidance.
\end{itemize}

\subsubsection{Integrating Emerging AI Technologies}

Emerging innovations offer new opportunities for pentesting:

\begin{itemize}
    \item \textbf{Advanced LLMs:} Explore multimodal LLMs for advance task processing and task-specific fine-tuning.
    \item \textbf{AI Agents for System Interaction:} Investigate agents capable of human-like system navigation \& interaction.
    \item \textbf{Localized Models:} Develop deployable, private AI tools to reduce dependency on external APIs.
\end{itemize}

These directions highlight the need for broader phase coverage, richer AI-human interaction, and tools that balance capability with privacy and scalability.

\section{Threats to Validity} \label{sec:threats}
\subsection{Threats to Validity}

Despite efforts to conduct a rigorous and impartial review, several threats may exist:

\begin{itemize}
    \item \textbf{Selection and Publication Bias:} Restricting to English, peer-reviewed sources, may exclude relevant non-English or unpublished work. Positive result bias in academic publishing can also skew findings.

    \item \textbf{Search Limitations:} Although multiple databases and keyword variants were used, relevant studies with alternative terminology may have been missed.

    \item \textbf{Context Dependence:} Included studies may reflect specific organizational or research settings, limiting generalizability to broader or industry-specific contexts where unpublished advancements may exist.

    \item \textbf{Heterogeneity and Synthesis Risk:} The diversity of pentesting practices makes standardization difficult; conclusions may oversimplify a complex and evolving landscape.
\end{itemize}


\section{Discussion \& Conclusion} \label{sec:conclusion}
This systematic literature review provides a comprehensive examination of the current state of AI-assisted penetration testing, highlighting both its emerging potential and the significant challenges that lie ahead.

\subsection{Key Findings}

The review yielded several core insights:

\begin{itemize}
    \item \textbf{Immature Landscape:} AI-assisted penetration testing remains in its early stages. Only one real-world application was identified, PenBox from the European Space Agency~\cite{Happe2023-il}.
    \item \textbf{Dominance of Reinforcement Learning:} Reinforcement Learning accounts for 77\% of the AI methodologies reviewed. Although this suggests strong promise, it also points to the need for further diversity in future work.
    \item \textbf{Phase Imbalance:} Most research targets the \textit{Discovery} and \textit{Exploitation} phases, leaving \textit{Preparation} and \textit{Reporting} phases comparatively underexplored.
\end{itemize}

\subsection{Technological Implications}
AI offers compelling advantages for penetration testing, particularly in enhancing speed and reducing manual workload, alongside challenges:

\begin{itemize}
    \item \textbf{Efficiency Gains:} AI can automate repetitive tasks and optimize attack strategies, helping human testers work more effectively.
    \item \textbf{Tooling Limitations:} AI tools remain narrow in scope, with no single solution offering broad testing capabilities.
\end{itemize}

\subsection{Practical Significance}

AI is not positioned to replace human pentesters, but to augment their capabilities:

\begin{itemize}
    \item \textbf{Assistive Role:} AI enhances the productivity of the tester by accelerating low value tasks and generating higher value insights.
    \item \textbf{Complexity Management:} As modern systems become more complex, AI offers scalable strategies to maintain a good security posture.
\end{itemize}

\subsection{Challenges and Constraints}

Despite its promise, AI-assisted pentesting faces important challenges:

\begin{itemize}
    \item \textbf{Technical Gaps:} Existing models lack generalizability across diverse systems and require retraining for new contexts.
    \item \textbf{Ethical Risks:} The automation of offensive security tasks raises concerns about misuse, transparency, and accountability.
\end{itemize}

\subsection{Conclusion}

AI stands at the cusp of revolutionizing penetration testing. While the current landscape is characterized by limited real-world applications and predominantly theoretical research, the potential is immense. As cyber threats continue to evolve in complexity, AI-assisted penetration testing represents a critical frontier for emerging cybersecurity research. AI can continue to enable pentesters to produce more thorough and advanced results in a more time-efficient manner, ultimately propelling this critical field forward. AI is not a replacement for penetration testers, but a tool to empower them.

\section{Data Availability} \label{sec:data}
The data underlying the findings of this study will be made accessible to the research community and can be found at \href{https://figshare.com/account/articles/29143250?file=54795278}{https://figshare.com/account/articles/29143250?file=54795278}. Currently, the data is shared privately but will be made publicly available upon acceptance of the manuscript.

\bibliographystyle{abbrv}
\bibliography{references}

\end{document}